\documentclass{JHEP3}%
\usepackage{amsmath}
\usepackage{mathbbol}
\usepackage{graphicx}
\usepackage{amssymb}
\usepackage{epsfig,multicol,bbm}
\usepackage{amsfonts}%
\providecommand{\U}[1]{\protect\rule{.1in}{.1in}}
\newcommand{\mytitle}{\title}
\newcommand{\myauthor}{\author}
\mytitle{Central exclusive production of longlived gluinos at the LHC}
\myauthor{P. J. Bussey$^1$, T.D. Coughlin$^2$, J.R. Forshaw$^2$ and A.D. Pilkington$^2$\\ $^1$Department of
Physics \& Astronomy, University of Glasgow, Glasgow G12 8QQ, UK.\\${}^2$ School of Physics \&
Astronomy, University of Manchester, Manchester M13 9PL, UK.}
\preprint{MAN/HEP/2006/10 \\ GLAS-PPE/2006-17}
\abstract{We examine the possibility of producing gluino pairs at the LHC via the exclusive reaction $pp \to p+\tilde{g}\tilde{g}+p$ in the case where the gluinos are long lived. Such long lived gluinos are possible if the scalar super-partners have large enough masses. We show that it may be possible to observe the gluinos via their conversion to $R$-hadron jets and measure their mass to better than 1\% accuracy for masses below 350 GeV with 300 fb$^{-1}$ of data.}
\begin{document}
\section{Introduction}

There is increasing interest in the possibility of instrumenting the LHC to
measure protons, scattered through very small angles, using detectors at a
distance down the beam pipe from the interaction point \cite{FP420,Totem}. A
primary motivation for installing such detectors is to study the process
$pp\rightarrow p+X+p$, where the protons and the central system, $X$, are
separated by large rapidity gaps. This process has already been studied
extensively in the literature and it is clear that it offers a unique
possibility to produce and explore new physics at the LHC
\cite{KMR,FP420,prospects,Forshaw}. In this paper, we consider the case in
which $X$ is a pair of long lived gluinos, forming either a bound state,
termed gluinonium~\cite{glu1,glu2,glu3,glu4}, or hadronising individually to
form colour singlet states, termed $R$-hadrons~\cite{Farrar}.

The possibility that the gluino may be long-lived is a hallmark of the
recently proposed `Split Supersymmetry' scenario~\cite{A&D, G&R}, though long
lived gluinos have been studied before, in the context of models in which the
gluino is the Lightest Supersymmetric Particle (LSP)~\cite{Farrar, Raby,
Baer}. In Split Supersymmetry the SUSY breaking scale, $m_{S}$, is large
($m_{S}\gg$1~TeV) and the scalar particles acquire masses at this scale. The
sfermions of the theory are protected by chiral symmetries and so can have
masses at the TeV scale as can one neutral Higgs boson whose mass is allowed
to be finely tuned. As a result the gluino can be long-lived on collider
timescales since it can only decay via the massive scalar particles.

The gluino lifetime is approximately given by (neglecting electroweak
corrections and possible decays to a Goldstino\footnote{The inclusion of
decays to a Goldstino can in some cases substantially decrease the gluino
lifetime~\cite{Gambino}.}):
\begin{equation}
\tau_{\tilde{g}}=\frac{\text{4~sec}}{N}\left(  \frac{m_{S}}{10^{9}~\text{GeV}%
}\right)  ^{4}\left(  \frac{1~\text{TeV}}{m_{\tilde{g}}}\right)  ^{5}\;,
\end{equation}
where $N$ depends upon $m_{S}$ and $m_{\tilde{g}}$ (and very weakly on
$\tan{\beta}$)~\cite{Gambino}, but is of order unity. If $m_{S}$ is unbounded
from above then so is the gluino lifetime. Cosmological considerations of long
lived gluinos in the early universe~\cite{wacker} can place upper bounds on
$\tau_{\tilde{g}}$, giving $\tau_{\tilde{g}}\lesssim100~\text{s}$ for
$m_{\tilde{g}}\gtrsim500~\text{GeV}$ and $\tau_{\tilde{g}}\lesssim10^{6}%
~$years for $m_{\tilde{g}}\lesssim500~\text{GeV}$. Clearly these constraints
do not rule out gluinos long-lived on collider timescales. In this paper we
consider the case where the gluinos are sufficiently long-lived that they do
not decay within the detector.\footnote{It means we do not consider the case
where the gluinos stop and subsequently decay within the
calorimeter~\cite{stopped}.}

Data from the Tevatron have been used to place the limit $m_{\tilde{g}}%
\gtrsim170~\text{GeV}$ on the mass of a long lived gluino~\cite{R-had6}, for
the case in which the gluino forms only neutral hadrons which remain neutral
as they pass through the detector. This limit is expected to rise to
$\simeq210$ GeV using Run II data \cite{R-had6}. We should stress that this is
a conservative limit, since it is anticipated that these hadrons will undergo
charge conversion reactions as they pass through the
detector~\cite{R-had7(Kraan1)}. In the most optimistic case, the Tevatron may
reach gluino masses of up to $\simeq430$ GeV if no signal is observed
\cite{R-had6}.

The structure of the paper is as follows: In Section~\ref{sec:Diff_Prod} we
give a description of the process $pp\rightarrow p+X+p$ and the model we shall
use to calculate cross-sections. In Section~\ref{sec:Gluinonium} we briefly
consider the production of the lowest lying colour singlet bound state of two
gluinos and show that the rate for bound state production is too low for it to
be interesting at the LHC. In Section~\ref{sec:R-hadrons} the signatures of
gluino hadronisation are discussed and the cross-section for open gluino
production is presented.

\section{Diffractive production with forward protons}

\label{sec:Diff_Prod} 
\begin{figure}[h]
\begin{center}
\centerline{\includegraphics[width=5cm]{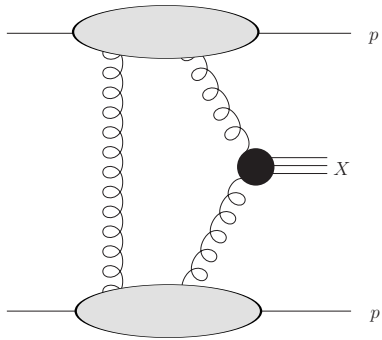}}
\end{center}
\caption{The central exclusive production process $pp\rightarrow pXp$.}%
\label{fig:pXp}%
\end{figure}

We calculate the cross-section using the model of Khoze,
Martin and Ryskin (KMR) \cite{KMR}, in which the process $pp\rightarrow p+X+p$
takes place as illustrated in Fig.\ref{fig:pXp}. The cross-section factorises
into a hard scattering part which represents the gluon fusion sub-process,
$d\hat{\sigma}(y,\hat{s})$, and an effective gluon luminosity, $d\mathcal{L}%
(\hat{s},y)/(dy~d\hat{s})$. As a result, we can write the cross-section for
producing any central system at rapidity $y$ and invariant mass $\hat{s}$ as
\begin{equation}
\frac{d\sigma}{dyd\hat{s}}=\frac{d\mathcal{L}(\hat{s},y)}{dyd\hat{s}}%
d\hat{\sigma}(y,\hat{s})\;. \label{eq:exclusive}%
\end{equation}
An overview of the details of the calculation has been given by one of us
\cite{Forshaw}. Here we just state the result:%
\begin{equation}
\hat{s}\frac{d\mathcal{L}(\hat{s},y)}{dyd\hat{s}}=\left[  \frac{\pi}{8b}%
{\displaystyle\int\limits^{M^{2}/4}}
\frac{dQ^{2}}{Q^{4}}f(x_{1},Q,M)f(x_{2},Q,M)\right]  ^{2} \label{eq:lumi}%
\end{equation}
where%
\begin{equation}
f(x,Q,M)=R_{g}Q^{2}\frac{\partial}{\partial Q^{2}}\left(  \sqrt{T(Q,M)}%
xg(x,Q^{2})\right)
\end{equation}
and
\begin{equation}
T(Q,M)=\exp\left(  -\int_{Q_{T}^{2}}^{M^{2}/4}\frac{dp_{T}^{2}}{p_{T}^{2}%
}\frac{\alpha_{s}(p_{T}^{2})}{2\pi}\int_{0}^{(1+2p_{T}/M)^{-1}}dz\;[zP_{gg}%
(z)+\sum_{q}P_{qg}(z)]\right)  .
\end{equation}
We take the diffractive slope parameter $b=4$ GeV$^{-2}$ and $g(x,Q^{2})$ is
the gluon distribution function of the proton. $R_{g}$ is a parameter which
corrects for the fact that we really need an off-diagonal gluon distribution.
We use the default value in the ExHuME Monte Carlo \cite{Exhume} which, at LHC
energies, corresponds to $R_{g}\approx1.2$ \cite{prospects}. Formally the $Q$
integral needs to be cut-off in order to avoid the pole in the running
coupling. In practice the integral is peaked well above $\Lambda_{\text{QCD}}$
and so the final results are insensitive to the value of the infra-red cutoff
(ExHuME takes a value of 800 MeV).

Since we are considering scattering of the protons through very small angles
there is an effective $J_{z}=0$ selection rule enforced upon the sub-process.
This means that the incoming gluons which fuse to produce the central system
$X$ should have equal helicities. We choose to normalise the luminosity
function in (\ref{eq:lumi}) such that an average over the incoming helicities
and the colour of the gluons must be included in the sub-process amplitude,
i.e.
\begin{equation}
\mathcal{M}=\frac{1}{(N_{C}^{2}-1)}\frac{1}{2}\left(  \mathcal{M}%
^{++}+\mathcal{M}^{--}\right)  \label{eq:col_pol_av}%
\end{equation}
and
\begin{equation}
d\hat{\sigma}(y,\hat{s})=\frac{1}{2\hat{s}}\left\vert \mathcal{M}\right\vert
^{2}d(\text{PS}).
\end{equation}

\section{Gluinonium}

\label{sec:Gluinonium} Close to threshold, the final state gluinos in the
process $pp\rightarrow p+\tilde{g}\tilde{g}+p$ may be produced in a colour
singlet bound-state configuration. Due to the Majorana nature of the gluinos
the sum of the spin and angular momentum quantum numbers, $S+L$, of the bound
state must be an even number~\cite{glu4}. Parity conservation then implies
that $^{3}\!P_{0}$ is the lowest accessible state. After production, the
gluinonium state would decay very rapidly to two gluons, which should then be
detected as a pair of jets.

Given the large mass of the gluinos, we can use a non-relativistic Coulomb
potential to determine the interaction between them, i.e.
\begin{equation}
V\left(  r\right)  =-\frac{3\alpha_{s}}{r}\;, \label{eq:potential}%
\end{equation}
where $r$ is the separation of the gluinos. The hard scattering cross-section
we require is related simply to the width for the decay of the bound state to
two gluons $\Gamma_{gg}$ via%
\begin{equation}
\frac{d\hat{\sigma}}{d\hat{s}}=2\frac{\pi^{2}}{M^{3}}\Gamma_{gg}\delta
(1-\hat{s}/M^{2}). \label{eq:sub-xsection}%
\end{equation}
This is the usual result for the production of a bound state of mass
$M\approx2m_{\tilde{g}}$ \cite{glu4} except that we need an additional factor
of 2 to account for the fact that this is to be used in the exclusive
cross-section formula (\ref{eq:exclusive}) above.

The width for the $P$-wave gluinonium decay to two gluons can be computed and
related to the derivative of the radial wavefunction at the origin, i.e.
\cite{prospects,novikov}%
\begin{equation}
\Gamma_{gg}=648\alpha_{s}^{2}(m_{\tilde{g}})\frac{\left\vert R^{\prime
}(0)\right\vert ^{2}}{M^{4}}=27\left(  \frac{3}{4}\right)  ^{5}~\alpha_{s}%
^{2}(m_{\tilde{g}})\alpha_{s}^{5}\!\left(  Q\right)  M
\end{equation}
where\footnote{$R(r)$ is normalized such that $\int dr~r^{2}R(r)^{2}=1$.}
\begin{equation}
R(r)=\frac{1}{\sqrt{24a_{0}^{5}}}r\text{e}^{-r/(2a_{0})}%
\end{equation}
and the Bohr radius is
\begin{equation}
a_{0}=\frac{2}{3}\frac{1}{m_{\tilde{g}}\alpha_{s}(Q)}\text{.}\nonumber
\end{equation}
We choose the scale $Q~$to be determined by the mean size of the gluinonium
state, i.e.%
\begin{equation}
Q=\frac{1}{\left\langle r\right\rangle }\approx\frac{1}{10a_{0}}.
\label{eq:Qscale}%
\end{equation}
This scale is substantially smaller than the value chosen in \cite{prospects}%
\textbf{\ }but ought to be more appropriate since the bound state is $P$-wave
rather than the more compact $S$-wave.

\begin{figure}[h]
\begin{center}
\centerline{\includegraphics[height=9cm]{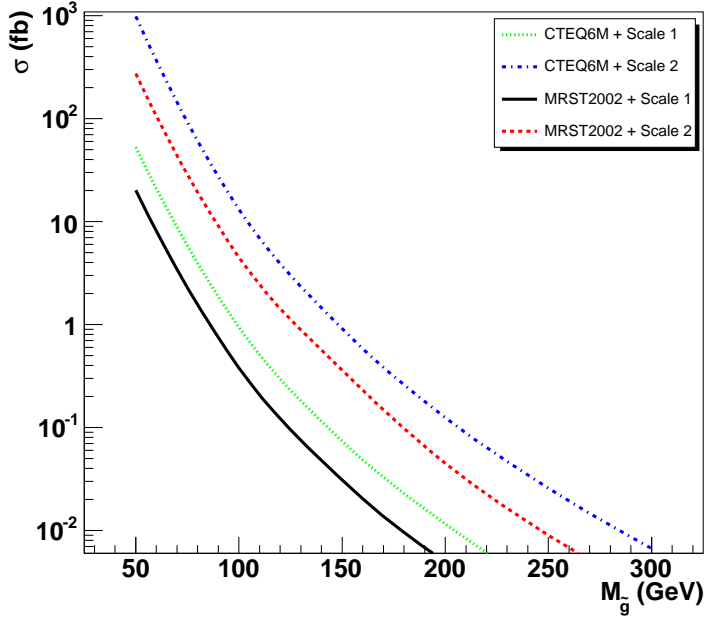}}
\end{center}
\caption{The total cross-section for gluinonium production. Shown are results
for the MRST2002nlo \cite{MRST} and CTEQ6m \cite{CTEQ} PDF sets with two
different choices of scale. Scale 1 is the choice $Q^{-1}=a_{0}$ and Scale 2
is the choice $Q^{-1}=10a_{0}$.}%
\label{fig:oniumXS}%
\end{figure}

In making our predictions we use the default soft survival factor of 3\%, as
appropriate to central production at the LHC \cite{gapsurvival}. Figure
\ref{fig:oniumXS} shows the total cross-section of central exclusive
gluinonium production as a function of the mass of the gluinonium. The results
are presented for two choices of the scale $Q$ and for two different choices
of parton distribution function (PDF).

Even before backgrounds are considered, this rate is small. At low luminosity
(10 fb$^{-1}$ per year) we can expect fewer than $\sim5$ events after one year
for gluino bound states of mass $\approx340$ GeV, which corresponds to the
current experimental lower limit. This number rises to $\sim45$ events at high
luminosity (100 fb$^{-1}$ per year) but this is still likely to be
insufficient given that the backgrounds from inclusive central production
($pp\rightarrow p+jj+X+p$) and exclusive dijet production ($pp\rightarrow
p+jj+p$) need also to be brought under control. We have used ExHuME v1.3.2
\cite{Exhume} and POMWIG~\cite{pomwig} to model the backgrounds and confirm
that they are indeed prohibitive.

\section{$R$-hadrons}

\label{sec:R-hadrons}

\subsection{Production and spectrum}

The gluinos may also be produced unbound. Sufficiently far from threshold, the
differential cross-section in the centre-of-mass (CM) frame given by
\cite{prospects}
\begin{equation}
\left(  \frac{d\hat{\sigma}}{d\Omega}\right)  _{\text{CM}}=\frac{9}{32}%
\frac{\alpha_{s}^{2}(\mu)m_{\tilde{g}}^{2}\beta_{\tilde{g}}^{3}}{(m_{\tilde
{g}}^{2}+|\boldsymbol{p}|^{2}\sin^{2}{\theta})^{2}}\;K,
\end{equation}
where $\beta_{\tilde{g}}$ and $|\boldsymbol{p}|$ are the CM\ speed and
momentum of the gluinos. Following the NLO\ calculations in
\cite{Beenakker:1996ch}, we evaluate the running coupling at scale $\mu
=\frac{1}{5}m_{\tilde{g}}$. $K$ is a threshold correction factor which we take
to be \cite{prospects,adel,Baer}%
\begin{equation}
K=\frac{Z_{g}}{1-\exp(-Z_{g})}\left(  1+\frac{Z_{g}^{2}}{4\pi^{2}}\right)
\label{eq:threshold}%
\end{equation}
where%
\begin{equation}
Z_{g}=\frac{3\pi\alpha_{s}(\beta_{\tilde{g}}m_{\tilde{g}})}{\beta_{\tilde{g}}%
}.
\end{equation}

The gluino is colour octet and in the case where it is long lived will
hadronise into bound states termed $R$-hadrons. These are colour neutral
states of a single gluino bound with gluons and/or quarks and anti-quarks.
There have been several studies of the spectrum of these states and their
interactions in the detector~\cite{Farrar, R-had1, Raby, Baer, R-had2, R-had3,
R-had4, R-had5, R-had6, R-had7(Kraan1), R-had8, R-had9(Kraan2),stopped}. It is
found that the states are nearly mass degenerate, with the $R$-mesons
($\tilde{g}\bar{q}q$) being slightly lighter than the lowest gluino-gluon
state, $R_{g}^{0}$,~\cite{R-had7(Kraan1), R-had1, R-had2, R-had3} and the
$R$-baryons ($\tilde{g}qqq$) being about 0.3~GeV heavier than
these~\cite{R-had2, R-had7(Kraan1)}. Only a small proportion of gluinos are
expected to form $R$-baryons and roughly half of the $R$-mesons formed will be
charged, with the rest neutral. The fraction forming $R_{g}$ states is unknown
and is therefore a free parameter.

\subsection{Interaction in detectors and triggering}

As a $R$-hadron passes through the detector, it will lose energy through
ionisation (if it is charged) and via hadronic interactions in the
calorimeters. $R$-mesons may be converted into $R$-baryons by scattering off
nucleons in the calorimeter, but $R$-baryon conversion to $R$-mesons is likely
to be negligible~\cite{R-had7(Kraan1)}. Also, the $R_{g}$ states are expected
to interact in the same way as a neutral $R$-meson, hence by the time the
$R$-hadron has passed through the calorimeter it may well be an $R$-baryon,
irrespective of how it started. Thus, 75\% of all $R$-hadrons are expected to
be charged after passing through the calorimeters (the ratio of charged to
neutral $R$-baryon states is 3:1)~\cite{R-had9(Kraan2)}. These events will
look like a muon within a jet (though more isolated than one resulting from a
heavy quark weak decay~\cite{R-had9(Kraan2)}) but with the particle arriving
significantly later at the muon chambers.

The difficulties in triggering on events involving $R$-hadrons at the ATLAS
experiment have been discussed in~\cite{R-had9(Kraan2)}. Essentially, the
level 1 triggers which make use of energy deposited in the calorimeters
(including the missing energy triggers) are useless because the $R$-hadrons
are expected to leave too little energy in the detector. The only other option
is to use the muon triggers. This presents its own problems however, since the
$R$-hadrons can be so delayed that they do not even arrive at the muon
chambers within the same bunch crossing. We therefore impose the following cuts:

\begin{itemize}
\item The pseudo-rapidity of each $R$-hadron should satisfy $|\eta|<2.4$. This
is the limit of the muon trigger at ATLAS.

\item The speed of the fastest of the two $R$-hadrons should lie in the range
$0.6<\beta<0.9$. This is the $R$-hadron which triggers the muon chambers. The
upper bound is chosen in order to eliminate all muon backgrounds whilst the
lower bound arises in order that the $R$-hadron arrive in time to trigger the event.

\item The speed of the slower $R$-hadron should lie in the range
$0.25<\beta<0.9$ where the lower bound is determined by the requirement that
the $R$-hadron is in the same event record as the faster $R$-hadron.
\end{itemize}

The ability of the $\beta>0.6$ cut to retain the complete event information is
questionable\footnote{We thank Thorsten Wengler for drawing our attention to
this.}. Although it is true that the $R$-hadron arrives in time to trigger the
event, there is the possibility that the readout of the muon trigger will
assign the event to the wrong bunch crossing. The wrong event would then be
read out from the other sub-detectors. This does not necessarily mean that the
information is unobtainable. For example, the TRT reads out in 75 ns time
slices which could be long enough to contain the relevant information.
Nevertheless more detailed studies are required before firm conclusions on the
utility of the inner detectors can be drawn. Fortunately, the forward detector
information should be available provided they are configured to take the
events either side of the accepted event. We therefore propose to trigger on
the fastest $R$-hadron and then check the pot readout to see if there are hits
in the forward detectors in either the same event or the previous one. In what
follows we shall assume that no information is available other than that
provided by the forward detectors and the muon chambers, fortunately this
should be sufficient to make an interesting measurement of the gluino mass.

The gluino mass can be determined given the mass ($M$) and rapidity ($y$) of
the central system, and the lab scattering angles of the two gluinos
($\theta_{1}$ and $\theta_{2}$) by solving the pair of equations
\begin{equation}
\frac{1}{\tan\theta_{1}}=\frac{\gamma}{k\sin\theta_{CM}}\left(  k\cos
\theta_{CM}+\frac{\beta M}{2}\right)  ,
\end{equation}%
\begin{equation}
\frac{1}{\tan\theta_{2}}=\frac{\gamma}{k\sin\theta_{CM}}\left(  -k\cos
\theta_{CM}+\frac{\beta M}{2}\right)
\end{equation}
where $\theta_{CM}$ is the CM\ scattering angle, $\beta=\tanh y$,
$\gamma=\cosh y$ and $k^{2}=\hat{s}/4-m_{\tilde{g}}^{2}$. The forward
detectors measure the momentum loss, $x_{i}$, of the protons and hence the
mass and rapidity of the central system can be reconstructed using
\begin{equation}
M^{2}=x_{1}x_{2}s\text{ \ \ and}\qquad y=\frac{1}{2}\ln{\frac{x_{1}}{x_{2}}}.
\end{equation}
The muon spectrometer can be used to determine the angles $\theta_{1}$ and
$\theta_{2}$ and so the gluino mass can be measured using only the muon and
forward detector information.

When quoting expected numbers of events we must also remember to multiply by a
factor $0.75\times0.6$. The 0.75 accounts for the fact that the $R$-hadron
must be charged and the 0.6 is an estimate of the muon trigger efficiency .
The efficiency of the muon trigger is usually greater than 85$\%$ at ATLAS
\cite{atlasTDR}. However, it is argued in \cite{R-had9(Kraan2)} that the
efficiency is lower in the case of $R$-hadrons because the $R$-hadron can
interact in the muon system and change to another $R$-hadron of different
charge.\begin{figure}[h]
\begin{center}
\centerline{\includegraphics[width=10cm]{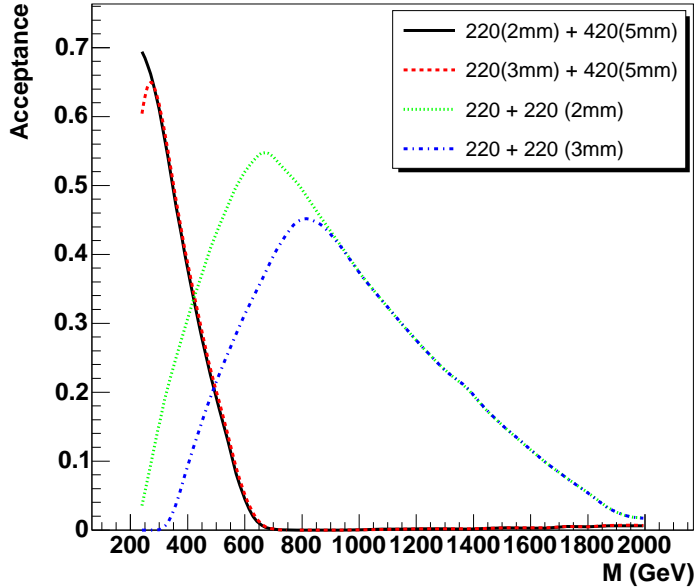}}
\end{center}
\caption{The acceptance of Roman Pots installed 5mm from the beam at 420m and
either 2mm or 3mm at 220m. }%
\label{acceptance}%
\end{figure}

We must also account for the acceptance of the forward detectors. At present
there are plans to put detectors at 220m \cite{Totem} and 420m \cite{FP420}
from the interaction point. Due to the relatively large masses we consider,
good acceptance for central masses in the range $300-1500$ GeV requires use of
at least one pot at 220m. The acceptances and resolutions for the measurements
in the Roman Pots have been calculated by means of the Monte Carlo program
FPTRACK which tracks the protons, generated using ExHuME, through an accurate
representation of the LHC beam line with the best available magnet optics. The
results for the acceptance are shown in Fig.\ref{acceptance}. Moving the 220m
pots closer to the beam, i.e. from from 3mm to 2mm, results in an increase in
acceptance in the mass range $200$ GeV$<$ $M<800$ GeV. Even in the most
conservative scenario with 420m pots at 5mm from the beam and 220m pots at 3mm
from the beam, the acceptance is more than 40\% up to central masses of 950 GeV.

There are two sources of background which need discussing. The first involves
processes in which the protons remain intact whilst the second arises as a
result of pile-up. The former comes mainly from the weak decays of exclusively
produced high transverse momentum $b$ and $t$ quarks, and it was found in
\cite{R-had9(Kraan2)} that these backgrounds can be effectively reduced by the
upper $\beta$ cut. Finally, there is a kinematic matching constraint, arising
from the requirement that the kinematics of the central system should match
the kinematics measured using in the pots, which will reduce the background
rate still further. We therefore believe that this type of background will be negligible.

The background from pile-up requires a different strategy for its removal. As
we shall see in the next section, gluino pair production is mainly interesting
when the LHC luminosity is high, i.e. $\sim 100$ fb$^{-1}$ per year. At
high luminosity, there are expected to be around 35 overlap events in each
bunch crossing and it is possible for two single diffractive events
($pp\rightarrow p+X$) to overlap with an inclusive hard scattering event
($pp\rightarrow X$). This could mimic the signal, with a proton from each of
the single diffractive events registering in the pots and the hard scattering
providing the candidate muons. However, this background can certainly be
reduced very substantially in a number of ways. Firstly, good timing from the
forward detectors can locate the candidate primary vertex to within 3mm
\cite{picosecond} and this can be compared to the position of the vertex
obtained from the central detector. Secondly, one can apply the kinematic
matching of the previous paragraph to ensure that the central system has
kinematics consistent with the measured protons. Finally, the muon candidates
will only pass the $\beta<0.9$ cut if the inclusive interaction occurs with a
time delay relative to the LHC clock. The spread of protons in a bunch at the
LHC is estimated to be $0.25-0.5$ ns. However, the minimum distance the muon
has to travel to trigger the event at high luminosity is $\simeq10$m, which
takes some 30~ns. Therefore, in order to pass the $\beta<0.9$ cut, the delay
of the interaction with respect to the LHC clock would have to be 3~ns. We
therefore conclude that backgrounds from this source will also be negligibly small.

Before moving on to present our numerical estimates, it is worth emphasising
that the possibility to measure central exclusive production even in the high
luminosity phase of the LHC is of much wider significance than the study
presented here. Potentially all processes hitherto examined in the literature
could benefit from the greatly enhanced statistics afforded if pile-up can be
brought under control. 

\subsection{Results}

\DOUBLEFIGURE [th] {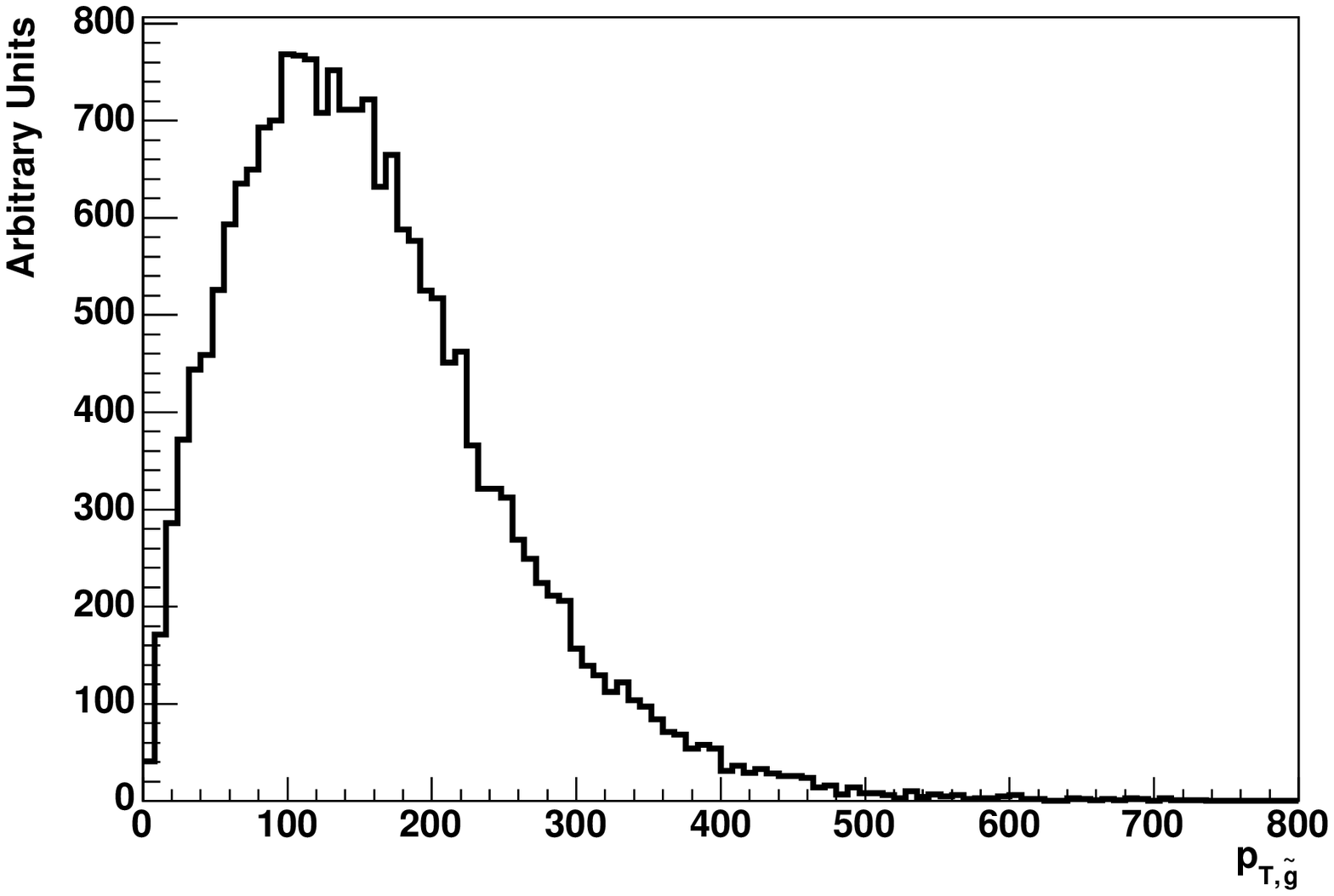,width=7cm}{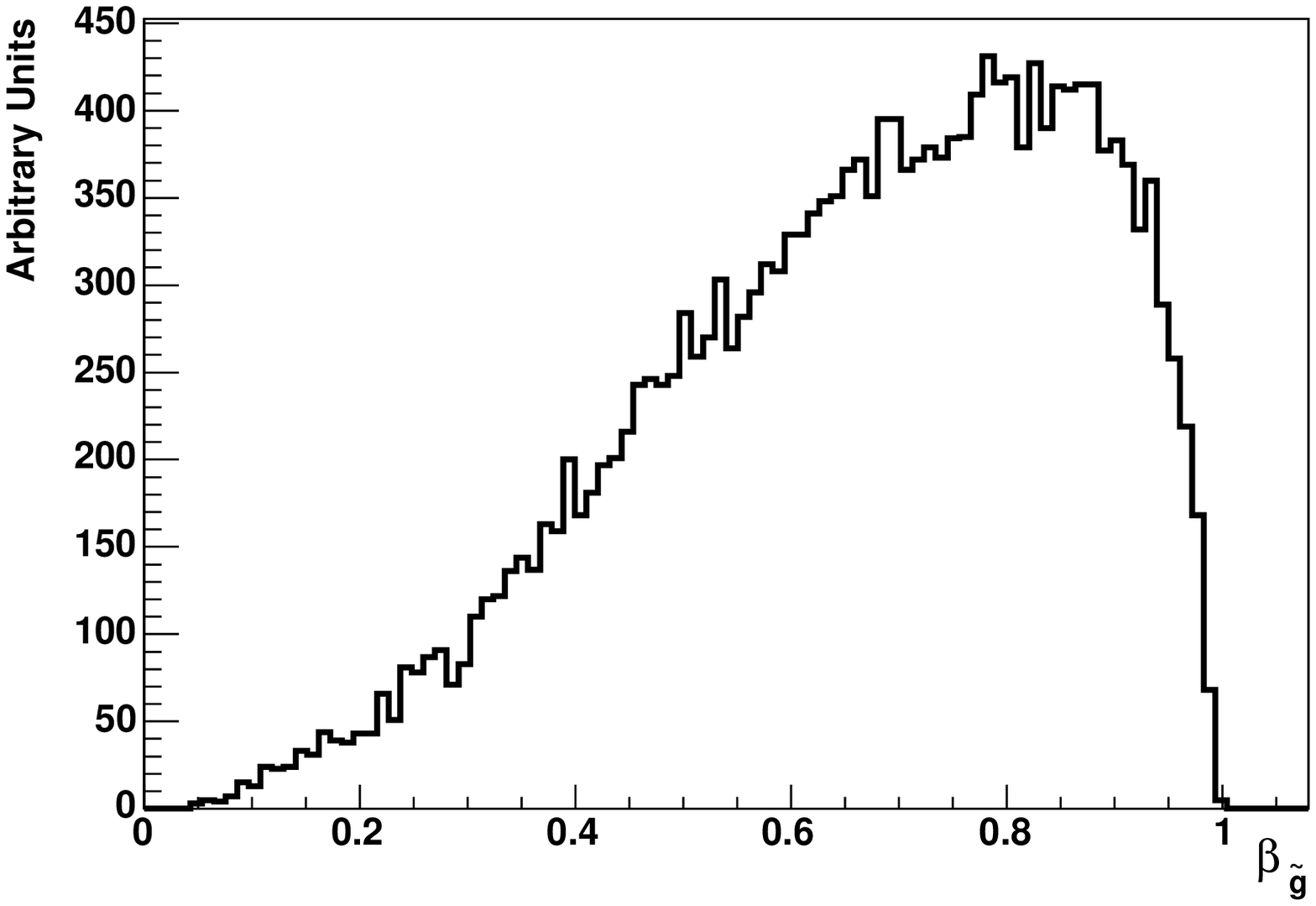,width=7cm} {The
$p_{T}$ distribution for pair producing 300 GeV gluinos.\label{ptgluino}} {The
$\beta$ distribution for pair producing 300 GeV gluinos.\label{betagluino}}

The cross-section for gluino pair production has been included in the ExHuME
Monte Carlo \cite{Exhume}. Figures \ref{ptgluino} and \ref{betagluino} show
the $p_{T}$ and $\beta$ spectra for 300 GeV gluinos. Clearly, the $\beta$ cuts
we use do not cut out too much signal. We do not need to worry about the
$p_{T}$ trigger cut because the muon trigger uses the radius of curvature of
the track in the muon spectrometer. As $R$-hadrons are much heavier than muons
they have a much larger radius of curvature for a given velocity and hence all
of them will pass the trigger cut. In Fig.\ref{gluinopairxs} we show the
cross-section for the central exclusive production of a pair of gluinos after
the cuts discussed in the previous section, but excluding the trigger
efficiency factor and the pot acceptance factor. We show our results for two
different PDFs. We also show the cross-section in the case that both
$R$-hadrons are required to make the level 1 trigger (i.e. the slower
$R$-hadron is also required to have $\beta>0.6$). In Fig.\ref{gluinothreshold}
we show the effect of the threshold correction given in (\ref{eq:threshold})
for our canonical $\beta$ cuts. Clearly the rate is not large. However, the
smallness of the background and the precision of the detectors means that only
very few events are needed in order to make an interesting measurement.

\DOUBLEFIGURE [t] {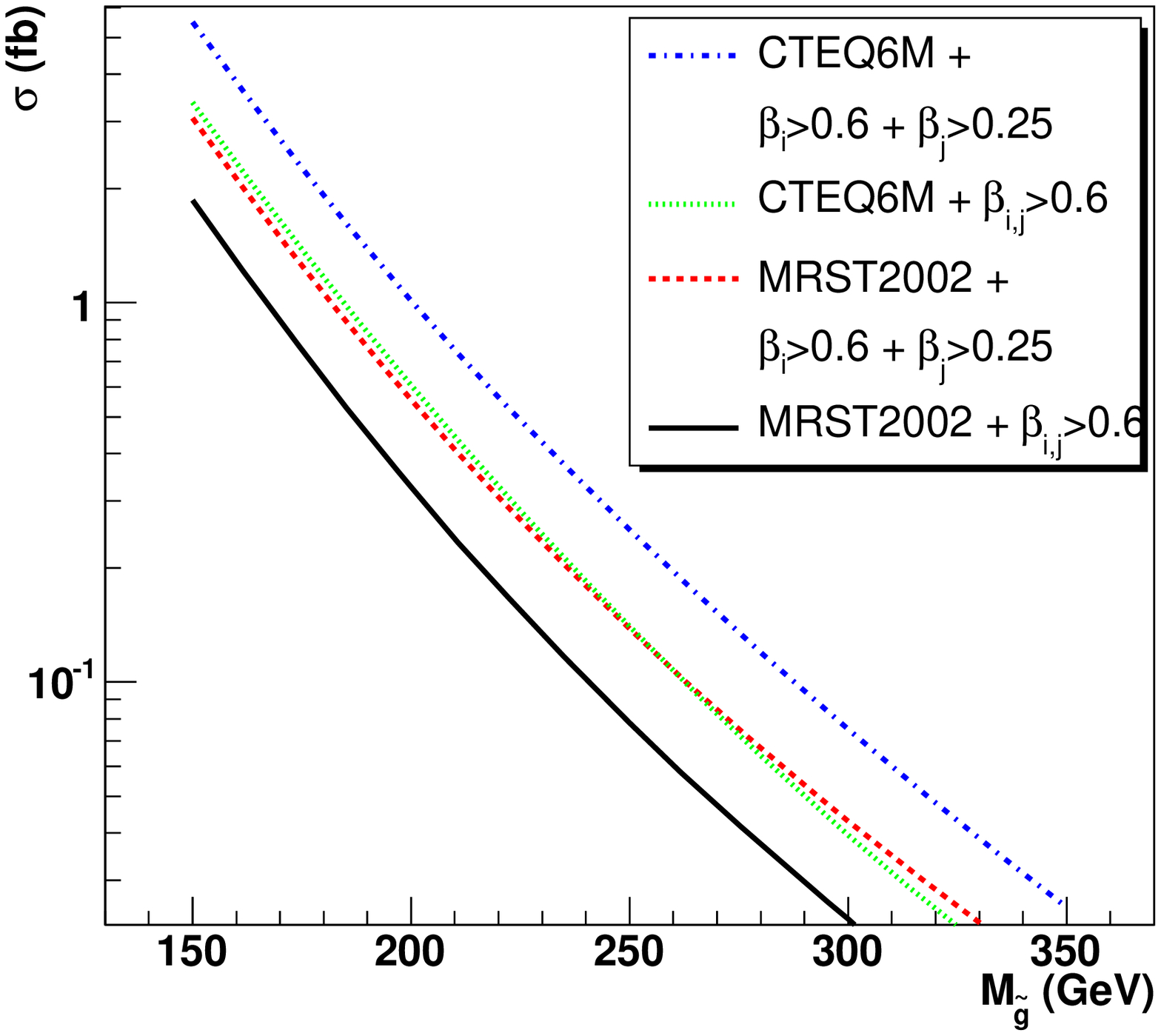,width=7cm}{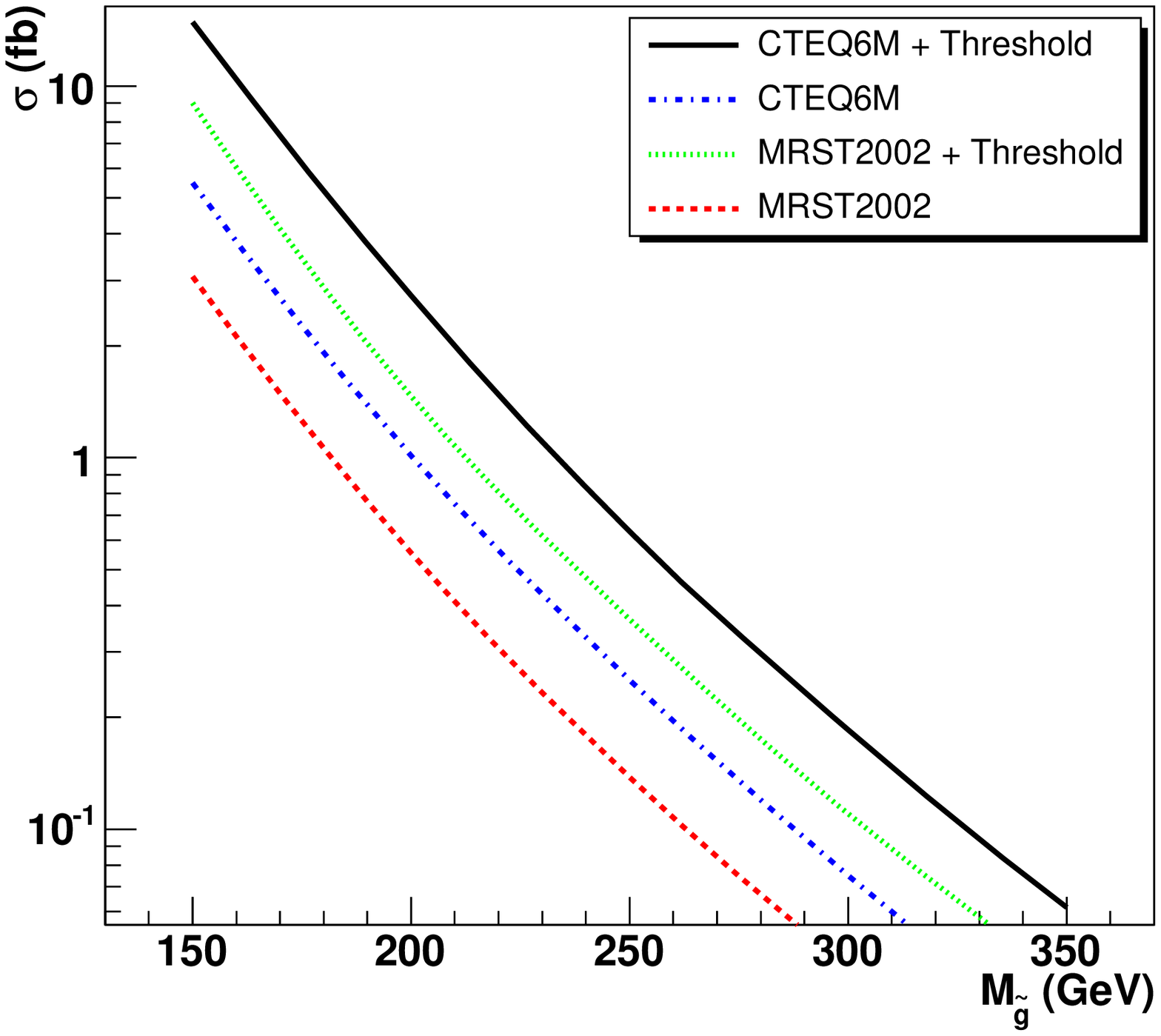,width=7cm}{The cross-section for exclusive gluino pair production for the MRST2002nlo and CTEQ6m PDF sets with two different choices of $\beta$ cut. \label{gluinopairxs}}{The effect of the threshold enhancement for MRST and CTEQ PDF's.\label{gluinothreshold}}

As we have seen, the gluino mass determination is possible using only the pots
and the muon detector. This is in contrast to the case where the gluinos are
produced inclusively (i.e. the protons break up). Inclusive production was
considered in \cite{R-had8}, where the reconstructed $R$-hadron mass was found
to be shifted from the true mass due to interactions of the $R$-hadron in the
calorimeter. The uncertainty of this shift was estimated to be 5 GeV for a 250
GeV gluino, with the systematic uncertainty becoming less important for larger
gluino masses. An analysis using the forward detectors thus complements that
of \cite{R-had8}, since it is best suited to making accurate measurements at
lower masses where the rate is higher.

\DOUBLEFIGURE [th] {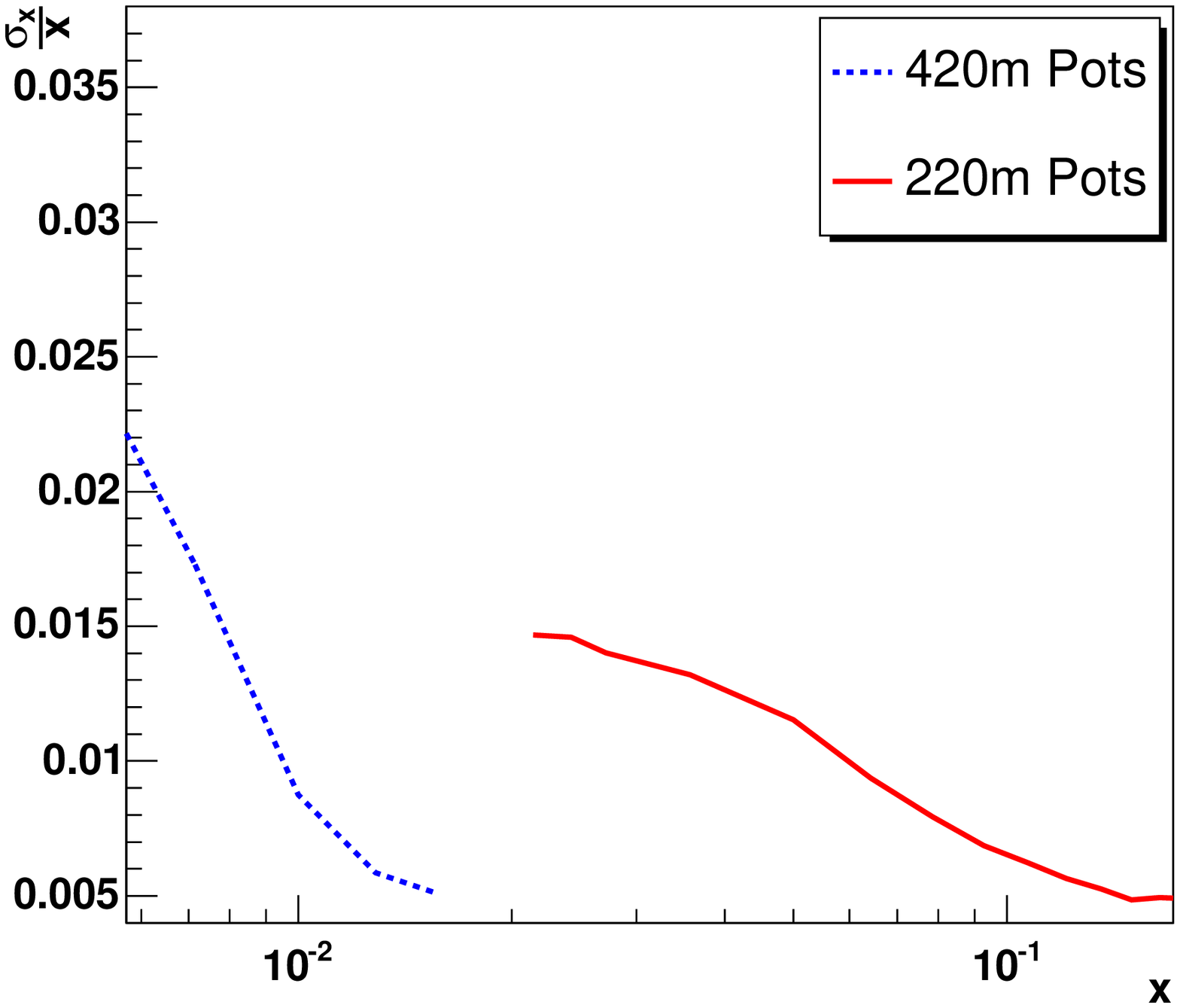,width=7cm}{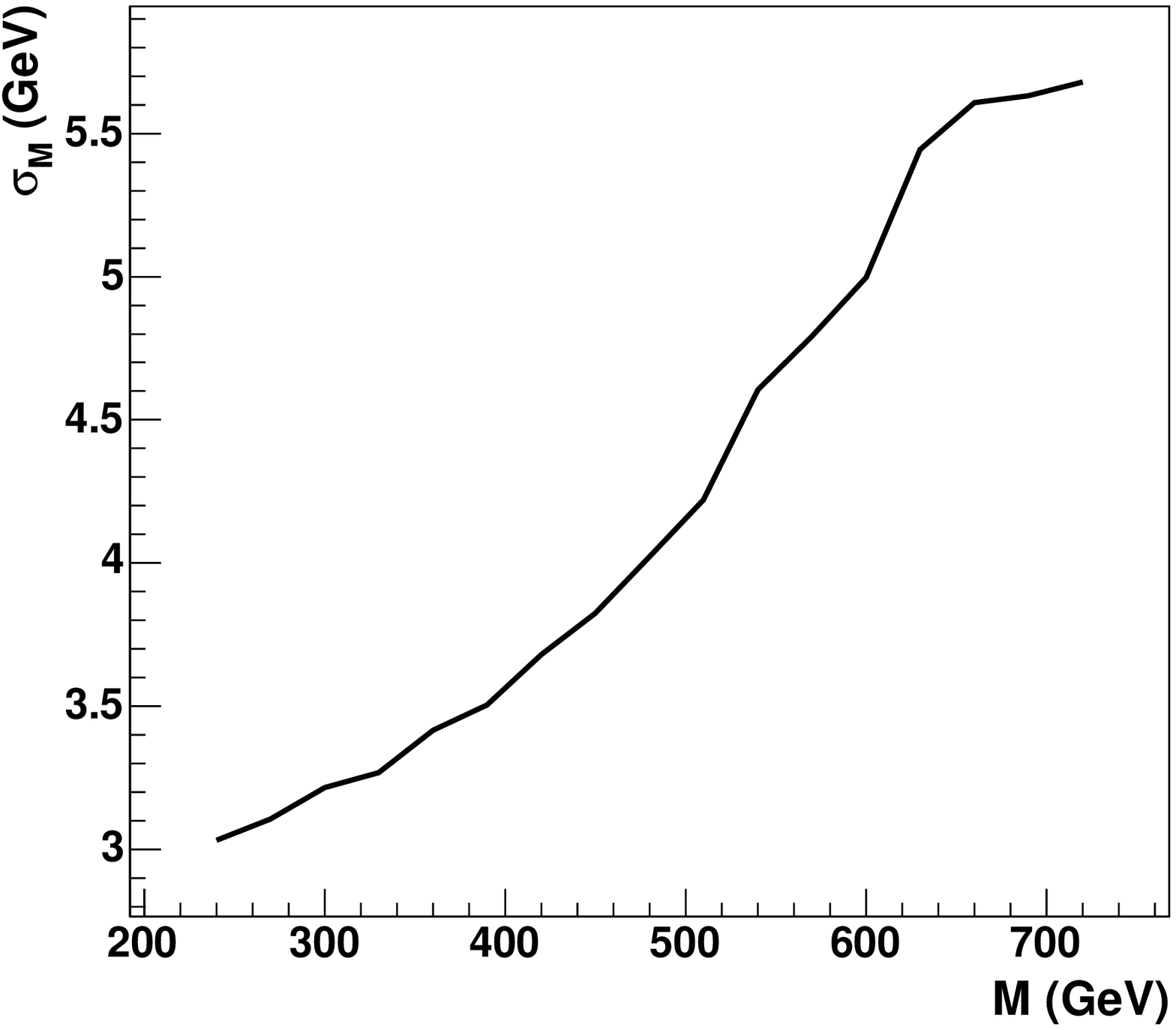,width=7cm} {The
resolution of the fractional momentum loss, $x$, of protons tagged in the
pots. \label{resolution}} {The resolution of the central mass $M$ (all
rapidities) \label{PR}}

Using FPTRACK, we obtain proton momentum resolutions as shown in
Fig.\ref{resolution}, from which the gluino mass resolution is calculated. The
resolution is dependent on the primary beam momentum spread, the beam spot
size and the angle/position measurement of the proton by the pots (1$\mu$rad
and 10$\mu$m). All that is required from the main detector itself are the
pseudo-rapidities of each of the gluinos. We smear the gluino
pseudo-rapidities using a gaussian distribution with $\sigma_{\eta}=0.0038$
which is our best estimate of the resolution that can be obtained using only
the muon detectors \cite{etares}. This uncertainty assumes that no inner
detector information will be available as a result of read-out problems. If it
transpires that some inner detector information is available, then this
resolution will obviously improve. The resulting gluino mass resolution, given
3 years of high luminosity running, is shown in Table
\ref{gluinomassresolution}. In particular we show the final error on the mass
measurement for $N$ events, where $N$ is determined using
Fig.\ref{gluinothreshold} after applying the pot acceptance ($0.4$) and
trigger efficiency ($0.75\times0.6$) factors. We see that it should be
possible to measure the gluino mass to an accuracy below 1\% up to gluino
masses of $\simeq350$ GeV.

\begin{table}[ptb]
\begin{center}%
\begin{tabular}
[c]{|c|c|c|c|}\hline
$m_{\tilde{g}}$ (GeV) & $\sigma_{m_{\tilde{g}}}$ (GeV) & $\frac{\sigma
_{m_{\tilde{g}}}}{\sqrt{N-1}}$ (GeV) & $N$\\\hline
200 & 2.31 & 0.19 & 145\\
250 & 2.97 & 0.50 & 35.0\\
300 & 3.50 & 1.10 & 10.2\\
320 & 3.61 & 1.54 & 6.5\\
350 & 3.87 & 2.45 & 3.5\\\hline
\end{tabular}
\end{center}
\caption{The gluino mass resolution as a function of the gluino mass. }%
\label{gluinomassresolution}%
\end{table}

We should comment upon the fact that the central masses we have been
considering are as large as 1 TeV. In this region, the theory we have used
should certainly be supplemented by the inclusion of quark exchange diagrams
and this should serve to increase the predicted rate. We therefore regard our
predictions as conservative.

\section{Conclusions}

We have shown that central exclusive production of gluino pairs
($pp\rightarrow p+\tilde{g}\tilde{g}+p$) could provide an accurate measurement
of the gluino mass at the LHC in the case that the gluinos are light enough
and sufficiently stable. Although detection of gluino bound states is not
viable, open gluino production could be detected provided the gluinos have
mass below $\simeq350$ GeV and their mass measured to an accuracy at the 1\%
level after 3 years of high luminosity running. Accurate mass determination is
a feature of having successfully detected and measured the scattered protons
using appropriately installed forward detectors during the high luminosity
phase of the LHC.

\section{Acknowledgements}

We should like to thank Alan Barr, Brian Cox, Stefano Rosati and Thorsten
Wengler for helpful discussions and Peter Richardson whose seminar initiated
the paper. This work was supported by the UK's Particle Physics and Astronomy
Research Council.


\begin{thebibliography}{99}                                                                                               %


\bibitem {FP420}FP420 project: M.G. Albrow et al, CERN-LHCC-2005-025 (2005).

\bibitem {Totem}TOTEM Collaboration: Technical Design Report, V. Berardi et
al, CERN-LHCC-2004-002 (2004).

\bibitem {prospects}V.A.~Khoze, A.D.~Martin and M.G.~Ryskin, Eur. Phys. J.
C23, 311 (2002).

\bibitem {KMR}V.~A.~Khoze, A.~D.~Martin and M.~G.~Ryskin, Phys.\ Lett.\ B401,
330 (1997).

V.~A.~Khoze, A.~D.~Martin and M.~G.~Ryskin, Eur.\ Phys.\ J.\ C14, 525 (2000).

\bibitem {Forshaw}J.R. Forshaw, in the proceedings of ``HERA and the LHC: A
workshop on the implications of HERA for LHC\ physics'', E-Print Archive: hep-ph/0508274.

\bibitem {glu1}K.~Cheung and W.~Keung, Phys. Rev. D71, 015015 (2005).

\bibitem {glu2}E.~Boudova-Thacker, V.~Kartvelishvili and A.~Small, Nucl. Phys.
Proc. Suppl. 133, 122 (2004).

\bibitem {glu3}E.~Chikovani, V.~Kartvelishvili, R.~Shanidze and G.~Shaw, Phys.
Rev. D53, 6653 (1996).

\bibitem {glu4}T.~Goldman and H.E.~Haber, Physica 15D, 181 (1985).

\bibitem {Farrar}G.R.~Farrar and P.~Fayet, Phys. Lett. D76, 575 (1978).

\bibitem {A&D}N.~Arkani-Hamed and S.~Dimopoulos, JHEP 0506, 073 (2005).

\bibitem {G&R}G.F.~Giudice and A.~Romanino, Nucl. Phys. B699, 65 (2004)
[Erratum-ibid. B706, 65 (2005)].

\bibitem {Raby}S.~Raby and K.~Tobe, Nucl. Phys. B539, 3 (1999).

\bibitem {Baer}H.~Baer, K.M.~Cheung and J.F.~Gunion, Phys. Rev. D59, 075002 (1999).

\bibitem {Gambino}P.~Gambino, G.F.~Giudice and P.~Slavich, Nucl. Phys. B726,
35 (2005).

\bibitem {wacker}A.~Arvanitaki, C.~Davis, P.W.~Graham, A.~Pierce and
J.G.~Wacker, Phys. Rev. D72, 075011 (2005).

\bibitem {stopped}A. Arvanitaki, S.Dimopoulos, A. Pierce, S. Rajendran and J.
Wacker, E-Print Archive: hep-ph/050642.

\bibitem {R-had6}J.L.~Hewett, B.~Lillie, M.~Masip and T.G.~Rizzo, JHEP 0409,
70 (2004).

\bibitem {R-had7(Kraan1)}A.C.~Kraan, Eur. Phys. J. C37, 91 (2004).

\bibitem {Exhume}J.~Monk and A.~Pilkington,
Comput.\ Phys.\ Commun.\ \textbf{175} (2006) 232. \verb|www.exhume-me.com|

\bibitem {novikov}V.A. Novikov et al, Phys. Rep. 41 (1978) 1.

\bibitem {gapsurvival}E.~Gotsman, E.~Levin, U.~Maor, E.~Naftali and
A.~Prygarin. E-Print Archive: hep-ph/0511060.

\bibitem {MRST}A.~D.~Martin, R.~G.~Roberts, W.~J.~Stirling and R.~S.~Thorne,
Eur.\ Phys.\ J.\ C28 (2003) 455

\bibitem {CTEQ}J.~Pumplin, D.~R.~Stump, J.~Huston, H.~L.~Lai, P.~Nadolsky and
W.~K.~Tung, JHEP 0207 (2002) 012

\bibitem {pomwig}B.~E.~Cox and J.~R.~Forshaw, Comput.\ Phys.\ Commun.\ 144
(2002) 104.\verb|www.pomwig.com|

\bibitem {Beenakker:1996ch}W.~Beenakker, R.~Hopker, M.~Spira and P.~M.~Zerwas,
Nucl.\ Phys.\ B492 (1997) 51.

\bibitem {adel}K. Adel and F.J. Yndurain, Phys.\ Rev.\ D52 (1995) 6577.

\bibitem {R-had1}M.S.~Chanowitz and S.R.~Sharpe, Phys. Lett. B126, 225 (1983).

\bibitem {R-had2}F.~Buccella, G.R.~Farrar and A.~Puliese, Phys. Lett. B153,
311 (1985).

\bibitem {R-had3}M.~Foster and C.~Michael [UKQCD Collaboration], Phys. Rev.
D59, 094509 (1999).

\bibitem {R-had4}G.~Karl and J.~Paton, Phys. Rev. D60, 034015 (1999).

\bibitem {R-had5}A.~Mafi and S.~Raby, Phys. Rev. D62, 035003 (2000).

\bibitem {R-had8}W.~Kilian, T.~Plehn, P.~Richardson and E.~Schmidt, Eur. Phys.
J. C39, 229 (2005).

\bibitem {R-had9(Kraan2)}A.C.~Kraan, J.B.~Hansen and P.~Nevski,
SN-ATLAS-2005-053, Nov 2005. E-Print Archive: hep-ex/0511014.

\bibitem {atlasTDR}ATLAS: Detector and physics performance technical design
report, Volume 1. CERN-LHCC--99-14.

\bibitem {picosecond}Talk given by A. Brandt at "Pico-Second Timing Hardware
Workshop" (2005), http://hep.uchicago.edu/workshops/2005-picosecond/

\bibitem {etares}Stefano Rosati, private communication.
\end{thebibliography}
\end{document}